# Modeling Heat Capacity of Supercritical Fluids: A Phonon Theory Approach

## Yu **Liu**\*, Chao **Liu**

Key laboratory of low-grade Energy Utilization Technologies and systems, Ministry of education, School of energy and power Engineering, Chongqing University, Chongqing ,400030, China

*Corresponding author: Tel: +86-023-65111826,

E-mail: Po_liuyu@163.com; liu_yu@cqu.edu.cn

# ABSTRACT

Recent research shows that liquids and dense supercritical fluids support high frequency shear waves. Here, we proposed a general heat capacity model of supercritical fluids using the latest theoretical findings (the liquid phonon theory). In this theory, heat capacity $c_v$ of supercritical fluids is only related to the infinite-frequency shear modulus, Debye relaxation time and minimal length of the longitudinal mode in supercritical fluids rather than the system-specific structure and interactions. The present model quantitatively explains the heat capacities in the liquid-like region and the gas-like region above the critical point. At the same time, the heat capacities of 6 supercritical fluids (including monatomic fluids Ar, Ne and molecular fluids $N_2$, CO, $CO_2$, $CH_4$) over a wide range of temperatures and pressures were calculated using the present model, yielding a high accuracy with no free-fitting parameters. This model has two advantages: (1) The contribution of intramolecular vibrations was considered, which makes the model applicable to molecular fluids and high temperature region; (2) This method starts from the quasi-harmonic Debye model. Thus, it does not require any adjustable parameters and thermal expansion coefficients. Furthermore, the present $c_v$ model of the rigid liquid-like supercritical region in this paper can be extended to liquid phase of molecular fluids.

**Key words**: Heat capacity; supercritical fluids; prediction; Debye model

# 1 Introduction

When the temperature and pressure of a fluid exceed its critical temperature and critical pressure, it enters a state known as the supercritical state, and the fluid itself is referred to as a supercritical fluid [1-3]. Supercritical fluid, is often hailed as the next-generation strategic technology for its wide utilization in energy, extraction, and cooling technologies [4-9]. Despite the widespread industrial application of supercritical fluids, our understanding of their thermodynamic properties, particularly heat capacity $c_v$, remains insufficient [10]. Researchers have employed phonon theory and the Boltzmann transport equation to study solids [11] and gases [12], deriving heat capacity $c_v=3R$ (Dulong-Petit Law) and $3R/2$ (kinetic theory of gases), respectively. However, when it comes to liquids and supercritical fluids, we found it is difficult to model them applying the same theories used for solids and gases [13]. Dealing with supercritical fluids faces a formidable challenge. Landau and Lifshitz emphasized that thermodynamic properties of dense fluids cannot be expressed as a general analytical form due to their significant molecule diffusion and strong interactions.

Currently, the main approach employed for modeling heat capacity of fluids is the Equation of State (EoS) [14-16]. In this method, Helmholtz energy was expressed as an empirical function of density ($\rho$) and temperature ($T$). Using EoS, one can derive the heat capacity. However, this method relies on numerous empirical parameters which lack clear physical significance and must be determined by experimental data of thermodynamic properties. Consequently, when attempting to extrapolate EoS to unexplored temperature and pressure region, the results are not guaranteed [17]. Molecular dynamics (MD) simulation plays a crucial role to investigate heat capacity [18]. By considering molecular interaction, it enables the prediction of various physical properties of fluids. MD provides a molecular-level understanding for heat capacity. This approach proves valuable when experimental properties of target fluids are scarce. Nevertheless, the performance of molecular dynamics simulation strongly depends on the potential function [19], which fails to accurately determine the heat capacity of fluids whose potential function is limited. Besides, the perturbation theory, the corresponding state principle and group contribution method were also used to calculate the heat capacity. Sheng et al [20] conducted a comprehensive review of available empirical models for heat capacity of fluids. Zaccone and co-workers [21-23] proposed a physically-based approach for heat capacity calculation using the vibrational density of states of liquids, this theory can be used to explain the typical monotonic decrease in heat capacity with temperature and provides a good single-parameter fitting to various experimental data for liquids. To sum up, the previous heat capacity models are practical but lack physical insight, suggesting that current understanding of supercritical fluids is incomplete [24]. In this work, we first analyzed the dynamic behavior of the supercritical fluids. Then, we proposed a physically-based heat capacity of supercritical fluids with no free-fitting parameters.

Unlike the traditional idea that physical properties are homogeneous above the critical point [25], new research [26] found the supercritical fluids exhibits two distinct states, namely a low-temperature, rigid liquid-like fluid and a high-temperature, non-rigid gas-like fluid, as shown in Figure 1. In fluid dynamics, the Frenkel line serves as the boundary in the pressure-temperature diagram. It delimits more liquid-like and more gas-like states of a supercritical fluid. Supercritical fluids on either side of this line are referred to as "liquid-like" and "gas-like" states, showing different behavior and structure. In rigid liquid-like supercritical fluids, transverse waves exist, resembling the behavior observed in solids. On the other hand, non-rigid gas-like fluids do not support shear modes, like gases. In the gas-like supercritical region, the system cannot maintain rigidity with any frequency, displaying gas-like characteristics.

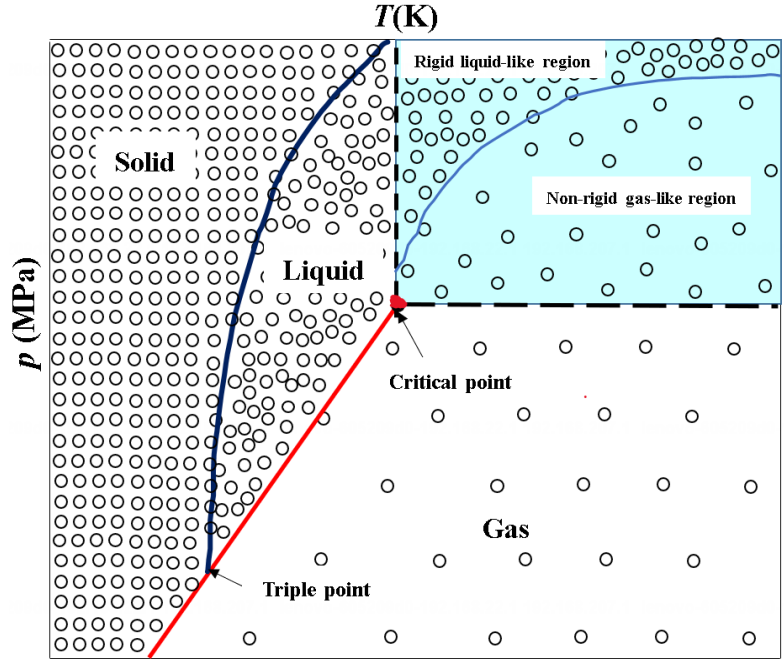

Figure 1. Phase diagram of solid, liquid, gas and supercritical matter

Heat capacity $c_v$ of a liquid-like supercritical fluid undergoes a significant decrease from approximately 3R to 2R as temperature increases [27-31]. This behavior can be explained quantitatively by the progressive loss of two shear waves with frequency $w > 1/\tau$. Frenkel's theory provides an explanation at the molecular level, suggesting that liquid-like supercritical fluids can be regarded as the solid structure on timescales shorter than the liquid relaxation time $\tau$ [32-34]. During these short durations or at frequencies greater than $w > 1/\tau$, solid-like shear waves can propagate in the supercritical fluid. Note that $\tau$ is the liquid relaxation time (the average time between continuous atomic jumps at a point in space), which is the vibration time of molecules at the equilibrium position. Neutron diffraction shows that Before the local structure is reorganized molecules in the liquid-like region can vibrate about 10 to 100 times near the original place, which is the approximate image of the molecular structure in liquids) [35].

When the relaxation time $\tau$ approaches $\tau_D$, the liquid-like supercritical fluid's ability to sustain transverse waves at any frequency diminishes [36]. At this point, the system's potential energy is solely dependent on the longitudinal modes, resulting in a total energy of 2RT and a heat capacity $c_v$ of 2R [10, 37]. Therefore, the decrease in $c_v$ from approximately 3R to 2R corresponds to the rigid liquid-like region, where short-duration and solid-like shear waves with high-frequency exist [38,39]. Note that the heat capacity of 2R is not coincidental but corresponds to a crossover occurring on the Frenkel line [40]. Simultaneously, the drop in $c_v$ represents the elimination of potential energy associated with transverse modes. As a result, supercritical fluids are no longer capable of sustaining shear waves at any frequency. Instead, the fluids transition into a new dynamic state resembling a non-rigid gas-like behavior, where molecule motion becomes purely collisional, like that in gases. Heat capacity $c_v$ continues to decline from 2R of crossover value to 3R/2 of ideal gas value as temperature increases.

Based on the fact that transverse waves with high frequency exist in liquids and dense supercritical fluids [26,41], Trachenko [13] firstly applied phonon excitations to calculate liquid heat capacity in 2008. Then Bolmatov and co-workers employed phonon theory, with some modifications, to derive the heat capacity of several supercritical fluids (e.g. Ar) [10]. Proctor [17] conducted a detailed analysis of the phonon theory of liquid thermodynamics. He concluded that the theory exhibits physically sensible parameters and the reasonable trend with density. This method circumvents the need for direct calculation of the interaction potential function between molecules. Besides, it eliminates the requirement for free-fitting parameters, exhibiting a high level of generality. However, the model proposed by Bolmatov et al overlooks the intramolecular vibrations inside the supercritical fluids. As a result, their model cannot be applied to substances with internal molecular structure in medium/high temperature region where the effect of intramolecular vibrations is obvious (e.g. $CO_2$ above room temperature, the intramolecular vibration's contribution to heat capacity is over 30%). Moreover, the model requires the thermal expansion coefficients of the target fluid, which also poses a limitation on its further application. In this paper, we developed an alternate model to improve these defects in previous liquid phonon theory. Notably, this model considers the contribution of intramolecular vibrations, enabling its application to molecular fluids and high temperatures. Additionally, the model originates from the quasi-harmonic Debye model, rendering it independent of thermal expansion coefficients of the target fluid.

The rest of this paper is arranged as below, Section 2 discusses heat capacities in liquid-like region and gas-like region separately. According to the latest findings above, a general heat capacity model of supercritical fluids was presented using the phonon theory. Section 3 calculates heat capacities of 6 common used fluids (Ar, Ne, $N_2$, CO, $CO_2$, $CH_4$) by means of the proposed model to test its performance. Section 4 explains the microscopic mechanism of heat capacity variation of monatomic and molecular supercritical fluids.

# 2 Development of the proposed model

**Low-temperature rigid liquid-like supercritical fluid.** Supercritical fluids involve two types of motion: phonon motion (molecular vibration around the equilibrium positions) and diffusional motion between adjacent positions, that is,

$$E = E_v + E_d \tag{1}$$

Here $E$ is the total energy. $E_v$ and $E_d$ are phonon and diffusional energy. The dynamic behavior of low-temperature rigid liquid-like supercritical fluids is similar to that of liquids, that is to say, supporting both longitudinal waves with all frequencies and shear waves with frequencies exceeding the Frenkel frequency $w_F$ (i.e. $w > w_F$, where $w_F = 2\pi/\tau = 2\pi G_\infty/\eta$, and the Maxwell relation $\tau = \eta/G_\infty$ is used to obtain liquid relaxation time $\tau$. $\eta$ and $G_\infty$ refer to viscosity and infinite-frequency shear modulus) [40].

$$E = K_l + P_l + K_s(w > w_F) + P_s(w > w_F) + K_d + P_d \tag{2}$$

Where $K$ and $P$ are kinetic and potential components of the energy. Subscript $l$, $s$ and $d$ represent longitudinal, shear and diffusion, respectively. According to the previous studies by Bolmatov et al [33][34], it has been found that $P_d$ in Eq.(2) is significantly smaller than the other terms, rendering it negligible. In that case, by combining the kinetic energy terms, the energy of the rigid liquid-like supercritical fluid can be expressed as below,

$$E = K + P_l + P_s(w > w_F) \tag{3}$$

Applying the energy equipartition theorem, we obtain $P_l = E_l/2$, $P_s(w > w_F) = E_s(w > w_F)/2$, and $K = K_l + K_s = E_l/2 + E_s/2$. Consequently, the total energy is then $E = E_l + E_s/2 + E_s(w > w_F)/2$. Since $E_s = E_s(w < w_F) + E_s(w > w_F)$, the total energy is given by,

$$E = E_l + E_s(w > w_F) + \frac{E_s(w < w_F)}{2} \tag{4}$$

Each term in the above formula can be calculated using the phonon free energy $F_{ph}$. Here, the quasi-harmonic Debye model [42] was introduced to obtain the energy term $E_{ph}$.

$$E_{ph} = F_{ph} - T \frac{dF_{ph}}{dT} \tag{5}$$

$$F_{ph} = \int \left[ \frac{1}{2}hw + k_B T \ln(1 - e^{-hw/k_B T}) \right] g(w) dw \tag{6}$$

For the longitudinal phonon energy $E_l$, the density of states $g(w)$ is $3N(w^2/w_D{}^3)$ while the integration range is from 0 to $w_D$. For the high-frequency shear phonon energy $E_s(w > w_F)$, the density of states $g(w)$ is $6N(w^2/w_D{}^3)$ while the integration range is from $w_F$ to $w_D$. For the low-frequency shear phonon energy $E_s(w < w_F)$, the phonon density of states $g(w)$ is $6N(w^2/w_D{}^3)$, integrating from 0 to $w_F$. Therefore, the phonon free energy can be expressed as follows,

$$F_l = \frac{1}{3}nRT \left[ \frac{9}{8}\frac{\theta_D}{T} + 3\ln(1 - e^{-\frac{\theta_D}{T}}) - D(\frac{\theta_D}{T}) \right] \tag{7}$$

$$F_s(w > w_F) = \frac{2}{3}nRT \left[ \frac{9}{8}\frac{\theta_D}{T} + 3\ln(1 - e^{-\frac{\theta_D}{T}}) - D(\frac{\theta_D}{T}) \right] - \frac{2}{3}nRT \left(\frac{w_F}{w_D}\right)^3 \left[ \frac{9}{8}\frac{hw_F}{k_B T} + 3\ln(1 - e^{-\frac{hw_F}{k_B T}}) - D(\frac{hw_F}{k_B T}) \right]$$

$$F_s(w < w_F) = \frac{2}{3}nRT \left(\frac{w_F}{w_D}\right)^3 \left[ \frac{9}{8}\frac{hw_F}{k_B T} + 3\ln(1 - e^{-\frac{hw_F}{k_B T}}) - D(\frac{hw_F}{k_B T}) \right]$$

$$D(x) = \frac{3}{x^3} \int_0^x \frac{z^3 dz}{\exp(z) - 1}$$

where $\theta_D$ represents the Debye temperature ($hw_D/k_B$). $D(x)$ stands for the Debye function.

$$c_v = dE/dT \tag{8}$$

For a monatomic fluid, its kinetic energy solely comprises translational kinetic energy ($c_v = c_{v,trans}$). However, the kinetic energy of molecular fluids consists of three contributions: the translational kinetic energy, the rotational kinetic energy and the vibrational kinetic energy. However, Eq.(4) and (8) only consider the translational kinetic energy, disregarding the rotational and vibrational kinetic energy components. When operating above room temperature, the rotational kinetic energy becomes fully excited, reaching its classical value. In the case of linear molecules, the contribution of rotational motion to heat capacity, $c_{v,rot}$, is equal to R. As for nonlinear molecules, $c_{v,rot}$ is 1.5R. Regarding to the contribution of vibrational kinetic energy to heat capacity, it still exhibits quantum features even above 1000 K due to its significant energy level spacing. Hence, within a broad temperature range, the contribution of vibration to the heat capacity, $c_{v,vib}$, is not in line with the classical value predicted by the energy equipartition theorem. Full excitation of vibrational energy occurs only at extremely high temperatures. In the commonly employed temperature range, vibrational energy demonstrates a quantum nature, where $hw \gg k_B T$ ($h$ represents Planck's constant, $w$ denotes frequency of the oscillator, $k_B$ signifies Boltzmann's constant, and $T$ stands for temperature). Consequently, the contribution of all vibrational degrees of freedom to heat capacity is determined by

$$c_{v,vib} = \sum_{i=1}^{n} R \left( \frac{\theta_{vi}}{T} \right)^2 e^{\theta_{vi}/T} \Big/ \left( e^{\theta_{vi}/T} - 1 \right)^2 \tag{9}$$

Where $n$ is the number of vibrational degrees of freedom. $\theta_{vi}$ is vibrational temperature in K and is defined as $\theta_{vi} = hw_0/k_B$. The vibrational temperature is commonly used to calculate the vibrational partition function. The vibrational temperatures of selected fluids are listed in Appendix.

Given the contribution of translational kinetic energy, rotation kinetic energy and vibration kinetic energy, heat capacity of rigid liquid-like supercritical fluids can be determined by summing these components.

**High temperature non-rigid gas-like fluid.** In the high-temperature non-rigid gas-like region, the supercritical fluid is unable to support shear waves at any frequencies while only longitudinal waves exist in the system. In this case, the total energy of the non-rigid gas-like supercritical fluids comprises kinetic energy and potential energy of longitudinal phonon with frequencies lower than $w_0$.

$$E = K + P_l (w < w_0) \tag{10}$$

Where $K$ represents the kinetic energy in the supercritical fluid. For monatomic fluids, only the translational kinetic energy ($E_{k,trans}$) is considered, which is given by $E_{k,trans} = 3RT/2$. However, for molecular fluids, both rotational kinetic energy and vibrational kinetic energy contribute to the total molecular kinetic energy. To calculate the upper limit of integration ($w_0$) in the system, the minimal wavelength of longitudinal modes, $\lambda$, present in the supercritical fluid was used. $w_0$ is determined by $w_0 = (2\pi/\lambda)c$, where $c$ refers to the speed of sound in the fluid.

In the high-temperature gas-like region of supercritical fluids, the shortest wavelength $\lambda$ is determined by the mean free path, which can be calculated using $\eta = \rho u \lambda / 3$, where $\rho$ is the density of fluids and $u$ is average velocity. Consequently, the mean free path can be utilized to determine the maximum frequency $w_0$ in a supercritical fluid.

Furthermore, based on the energy equipartition theorem, potential energy of longitudinal phonon with frequencies lower than $w_0$ can be written as $P_l(w<w_0) = E_l(w<w_0)/2$. Here, $E_l$ represents the longitudinal wave energy, which can be calculated based on the longitudinal phonon free energy (Eq.(6)). The density of states is determined using the Debye density of states, given by $g(w) = 3N(w^2/w_D{}^3)$. By integrating the formula for longitudinal phonon free energy from 0 to $w_0$, the phonon free energy is therefore,

$$F_l = \frac{1}{3}nRT\left(\frac{w_0}{w_D}\right)^3 \left[\frac{9}{8}\frac{\theta_0}{T} + 3\ln(1 - e^{-\frac{\theta_0}{T}}) - D(\frac{\theta_0}{T})\right] \tag{11}$$

Substituting Eq.(11) into Eq.(5), the longitudinal wave potential energy with a frequency lower than $w_0$ $P_l(w<w_0)$ can be obtained. Here $\theta_0 = hw_0/k_B$. Combined with Eq.(10), the energy of the non-rigid gas-like supercritical fluid can be determined. Deriving the energy allows to obtain heat capacity $c_v$ of the gas-like region,

$$c_v = \frac{d(K + P_l(w < w_0))}{dT} = c_{v,K} + c_{v,P} \tag{12}$$

Note that $c_{v,K}$ includes three contributions: translational kinetic energy component $c_{v,\text{trans}}$, rotational kinetic energy component $c_{v,\text{rot}}$ and vibrational kinetic energy component $c_{v,\text{vib}}$. As mentioned before, for monatomic fluids, its total heat capacity $c_v$ only counts the first part $c_{v,\text{trans}}$. However, for molecular fluids with internal structures, both the latter two components, $c_{v,\text{rot}}$ and $c_{v,\text{vib}}$, need to be taken into consideration. At room temperature and above, the rotational kinetic energy component of heat capacity $c_{v,\text{rot}}$ is fully excited and reaches its classical value. For linear molecules, $c_{v,\text{vib}} = R$, while for nonlinear molecules, $c_{v,\text{vib}} = 1.5R$. Regarding the contribution of vibrational kinetic energy $c_{v,\text{vib}}$, it can be calculated using formula (9). Then, heat capacity of non-rigid gas-like supercritical fluids can be obtained.

**Extrapolation Performance.** By examining the free energy expression of the rigid liquid-like region, Debye model and Dulong-Petit Law can be obtained by differentiating the free energy equation (Eq.(7)) along the isotherm, assuming that the Debye temperature $\theta_D$ is solely dependent on the density. Similarly, by deriving the free energy model of the non-rigid gas-like supercritical fluids (Eq.(10)), heat capacity of the gas, that is, $c_v=3R/2$, can be obtained. The remarkable extrapolation demonstrates that the present model has a solid physical foundation.

# 3 Heat Capacity of Selected Supercritical Fluids

Based on the $c_v$ formulation mentioned above, heat capacity of six supercritical fluids (Ar, Ne, $N_2$, CO, $CO_2$, $CH_4$) were calculated in the temperature range from 58 to 780 K and pressures up to 913 MPa. The calculation process is outlined below,

1, To obtain the liquid relaxation time and the shortest wavelength of the longitudinal mode based on the fluid viscosity.

2, To substitute the obtained relaxation time and shortest wavelength into Eq.(5), (7) and (11) to derive the phonon free energy and energy.

3, To calculate the derivative of energy to obtain the heat capacity. This step does not consider the intra-molecular vibration and rotation.

4, To obtain the total heat capacity of the supercritical fluid by adding the intra-molecular rotational kinetic energy component $c_{v,rot}$ and intra-molecular vibration kinetic energy component $c_{v,vib}$. After following these steps, the calculated heat capacities were then compared with the experimental data from NIST REFPROP [43].

It can be observed that the present phonon theory calculations show excellent agreement with the reference values, see Figures 2 to 7. Taking Argon as an example, the Debye temperature ($\theta_D$) is taken from the corresponding value of solid Ar (about 90 K) while the infinite-frequency shear modulus ($G_\infty$) is approximately 0.2 GPa during the calculation process. The physical parameters are close to those stated in Bolmatov's work [13]. Note that phonon theory does not account for the abnormal variation of heat capacities in the near-critical region.

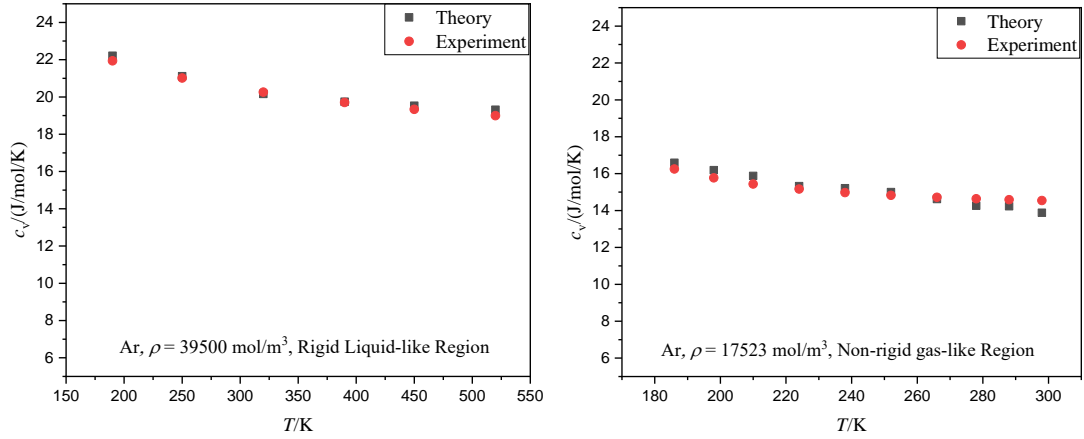

Figure 2. Calculated and Experimental $c_v$ of supercritical Ar.

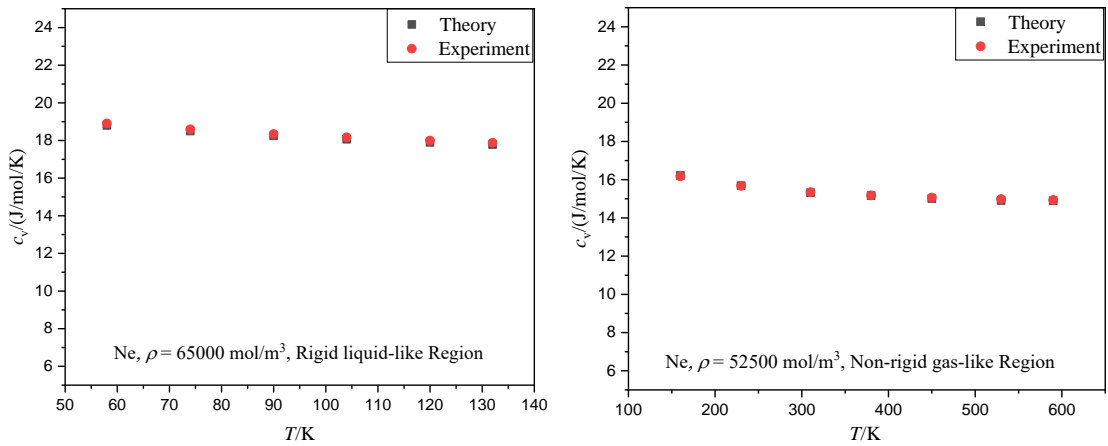

Figure 3. Calculated and Experimental $c_v$ of supercritical Ne.

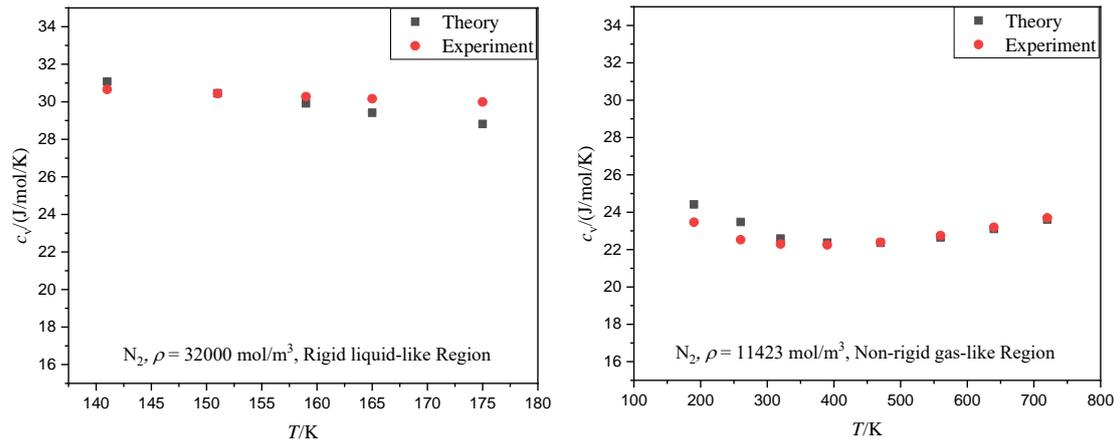

Figure 4. Calculated and Experimental $c_v$ of supercritical $N_2$.

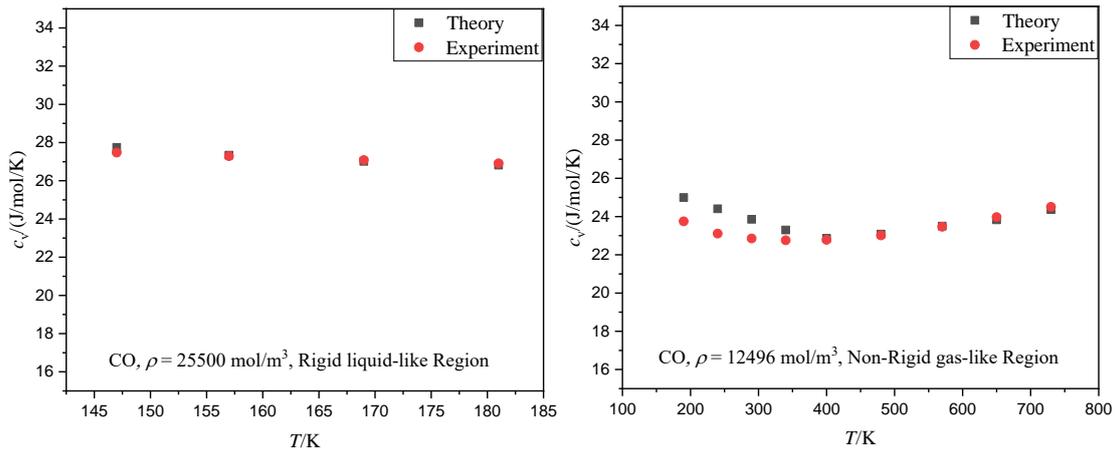

Figure 5. Calculated and Experimental $c_v$ of supercritical CO.

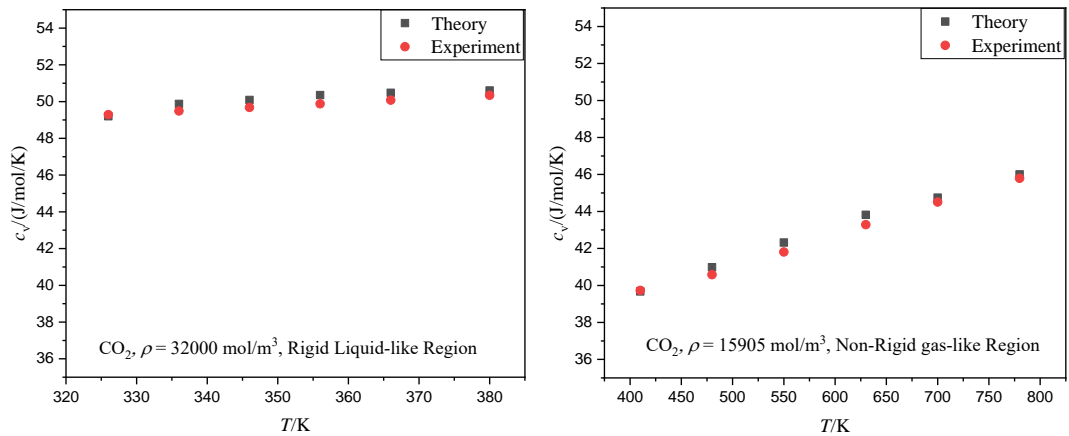

Figure 6. Calculated and Experimental $c_v$ of supercritical $CO_2$.

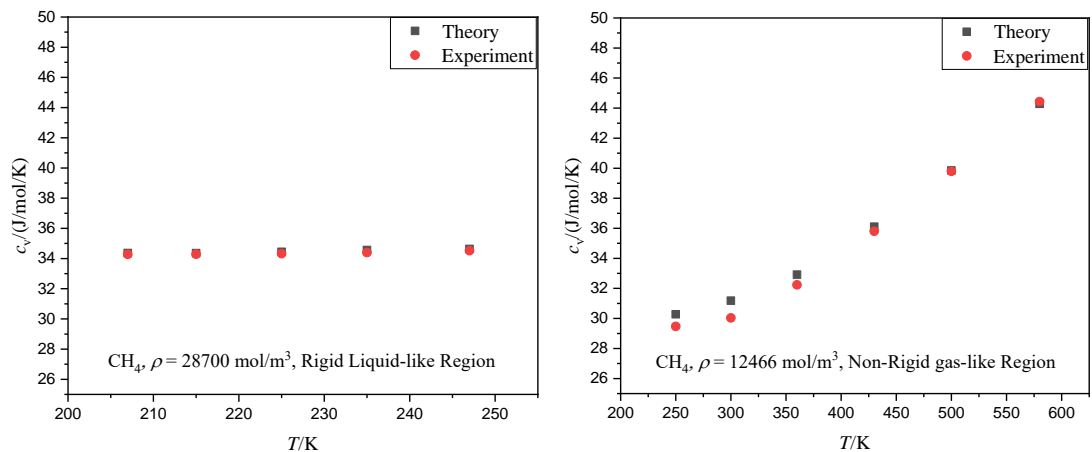

Figure 7. Calculated and Experimental $c_v$ of supercritical $CH_4$.

# 3 Molecular-Level Understanding of Heat Capacity of Supercritical Fluids

In addition, the variation of heat capacity for supercritical fluids was explained using the phonon theory present in this work.

**For monatomic fluids,** in the liquid-like region, both the kinetic energy $K$ and longitudinal wave potential energy $P_l$ in Eq. (3) increase rapidly with temperature (i.e. the slope $d(K+P_l)/dT$ gradually increases). However, as the temperature rises, the high-frequency shear wave potential energy $P_s(w>w_F)$ increases slowly and even decreases (i.e. the slope $dP_s(w>w_F)/dT$ gradually decreases and even becomes negative). While these two factors work together, the trend of decreasing slope predominates. Since the slope corresponds to the heat capacity, heat capacity of liquid-like supercritical fluids decreases with increasing temperature. In the gas-like region, the kinetic energy term $K$ in Eq. (10) increases with temperature (slope $dK/dT=1.5R$), while the low-frequency longitudinal wave potential energy $P_l(w<w_0)$ decreases with temperature (slope $dP_l(w<w_0)/dT$ is negative). Again, the two factors work together, and the trend of decreasing slope dominates. As a result, heat capacity of gas-like supercritical fluids decreases with temperature and eventually reaches 1.5R.

**For molecular fluids**, many factors affect their behavior, including intermolecular and intramolecular energy. Intermolecular energy refers to the interactions between molecules, while intramolecular energy involves molecular rotations and vibrations. At room temperature and above, intramolecular rotation reaches its classical value ($c_{v,rot}$ = R or 1.5R). On the other hand, intramolecular vibration energy is quantized, meaning it increases gradually with temperature. For molecular fluids, the slope of intermolecular energy versus temperature shows a downward trend, as discussed in the last paragraph. In contrast, the slope of intramolecular energy versus temperature exhibits an upward trend. These two opposing contributions compete and can lead to heat capacity showing either a downward or upward trend. Notably, the higher the temperature, the more obvious this upward trend in heat capacity. Taking $CO_2$ as an example, its supercritical heat capacity increases as temperature, as shown in Figure 6.

# 4 Conclusions

Recent research shows that the supercritical area was divided into two distinct regions, namely the liquid-like region and the gas-like region. In the liquid-like region, the supercritical fluid comprises longitudinal phonons and high-frequency shear phonons. As the temperature increases, the shear phonons gradually diminish, leading the fluid to enter the gas-like region, where only longitudinal waves exist. Based on the phonon theory, the phonon free energy was calculated with a general analytical form. Then the energy of the supercritical fluid was determined using free energy. Subsequently, the heat capacity of the supercritical fluid was derived by computing the temperature derivative of the system energy. Using the proposed model, the heat capacities for six fluids, including both monatomic and molecular fluids, were obtained. A comparison was made between the calculated heat capacities and reference data, showing a good agreement. The proposed supercritical heat capacity model eliminates the need for free-fitting parameters and thermal expansion coefficients. Moreover, it considers the influence of intramolecular kinetic energy and explains the variation in heat capacity of molecular fluids based on physical mechanisms.

## Appendix

Vibrational temperatures of selected fluids [44]

| Fluid | $\theta_v$,K | Fluid | $\theta_{v1}$,K | $\theta_{v2}$,K | $\theta_{v3}$,K | $\theta_{v4}$,K |
|-------|------|-------|-------|-------|-------|-------|
| $N_2$ | 3390 | $CH_4$ | 1870(3) | 2170(2) | 4320(3) | 4400 |
| CO | 3120 | $CO_2$ | 960(2) | 1330 | 3380 | - |
| NO | 2740 | $N_2O$ | 844(2) | 1842 | 3195 | - |
| $O_2$ | 2270 | $SO_2$ | 750 | 1650 | 1950 | - |
| $Cl_2$ | 807 | $NH_3$ | 1357 | 2336(2) | 4176 | 4776(2) |

## Nomenclature

| | |
|---|---|
| $c_v$ | isochoric heat capacity |
| $E$ | energy |
| $F$ | free energy |
| $G_\infty$ | infinite-frequency shear modulus (GPa) |
| $h$ | reduced Planck constant |
| $k_B$ | Boltzmann constant |
| $K$ | kinetic energy |
| $p$ | pressure (MPa) |
| $P$ | potential energy |
| $T$ | temperature (K) |
| Greek symbols | |
| $\eta$ | viscosity (Pa*s) |
| $\rho$ | molar density mol/m$^3$ |
| $\tau$ | liquid relaxation time (ps) |
| $w$ | frequency (Hz) |
| Subscripts | |
| d | diffusion |
| D | Debye |
| F | Frenkel |
| l | longitudinal wave |
| ph | phonon |
| rot | rotation |
| s | shear waves |
| v | vibration |

# AUTHOR INFORMATION

**Corresponding Author**

E-mail: Po_liuyu@163.com; liu_yu@cqu.edu.cn

Tel: +86-15619317108,

# Notes

The authors declare no competing financial interest.

# Acknowledgement

We acknowledge the support of the National Natural Science Foundation of China (Grant No. 52106218), China Postdoctoral Science Foundation funded project (Grant No. 2021M693713).